\documentclass[12pt]{article}
\def\bib{\bibitem}
\def\eg{{\em e.g.}}
\def\be{\begin{equation}}
\def\ee{\end{equation}}
\def\barr{\begin{array}}
\def\earr{\end{array}}

\def\gev{\: {\rm GeV} }
\def\tev{\: {\rm TeV} }

\def\ptsl{p_T \hspace{-1.1em}/\;}

\def\dis{\displaystyle}
\def\sh{\hat s}
\def\th{\hat t}
\def\uh{\hat u}

\begin{document}

\setlength{\unitlength}{0.1bp}


\vspace*{-0in}
\renewcommand{\thefootnote}{\fnsymbol{footnote}}
\setcounter{page}{0}
\thispagestyle{empty}
\begin{flushright}
CERN-TH/97-139  \\
TIFR-TH/97-28  \\
{\large \tt hep-ph/9706536} \\
\end{flushright}
\vskip 50pt
\begin{center}
{\Large \bf \boldmath $W + 1$ jet to $W + 0$ jet Ratio at the
                      Tevatron: \\[1.5ex]
               A Hint of New Physics? }\\
\vspace{8mm}
\vspace{15pt}
{\bf Debajyoti Choudhury\footnote{debchou@mail.cern.ch}, 
Sreerup Raychaudhuri\footnote{sreerup@mail.cern.ch}}\\
\vspace{6pt}
{\it Theory Division, CERN, CH-1211, Geneva 23, Switzerland.}\\
\vspace{13pt}
and \\
\vspace{8pt}
{\bf K. Sridhar\footnote{sridhar@theory.tifr.res.in}}\\
\vspace{6pt}
{\it Theory Group, Tata Institute of Fundamental Research,\\
Homi Bhabha Road, Bombay 400 005, India.}

\vspace{70pt}
{\bf ABSTRACT}
\end{center}
We interpret the reported disagreement between the measured ratio, 
$R_{10}$, of the
$W+1$ jet cross-section to the $W+0$ jet cross-section at the Tevatron 
and the Standard Model (SM) prediction, 
as the effect of interactions mediated by a colour-octet analogue 
of the $W$ boson, the $W_8$. The presence of a $W_8$ with mass 
${\cal O}(300)$ GeV, and with couplings to quarks and gluons of the order of 
electroweak strength, allows the observations of the D0 collaboration to 
be reproduced quite accurately. Such an interaction is not in contradiction
with the present CDF data on $W^+W^-$ or dijet production, though higher
luminosities may reveal measurable effects. \\

\vspace{0.2in}

\noindent
CERN-TH/97-139\\
June 1997
\vfill
\newpage
\noindent

\setcounter{footnote}{0}
\renewcommand{\thefootnote}{\arabic{footnote}}
\vfill
\clearpage
\pagestyle{plain}

Among the various physics programmes at hadron colliders, studies of the 
production of $W+$ jet(s) have attracted considerable interest since this
is an important testing ground for next-to-leading order (NLO) QCD 
corrections to the Drell-Yan process~\cite{NLO_Wg}. One particular observable 
in this context is the ratio, $R_{10}$, of the $W+1$ jet cross-section to
the $W+0$ jet cross-section. As several systematic effects tend to cancel 
in the ratio, $R_{10}$ seems likely to yield
a clean measurement of the strong coupling constant $\alpha_s$.
It was 
with this motivation that the ratio was originally measured by the UA1 
and UA2 collaborations~\cite{UA12} at the CERN $Sp \bar p S$ collider
at a centre-of-mass energy of 630 GeV. At the Tevatron, operating at the 
much higher centre-of-mass energy of 1.8~TeV, the D0 collaboration has 
carried out a more precise determination~\cite{D0_Wg_new}
of this ratio and finds that, in fact, the 
measured quantity has very little dependence on $\alpha_s$. With hindsight,
this paradox 
can be attributed to the fact that the variation with $\alpha_s$ of the 
parton-level
cross-section for $W+$ jet(s) is compensated by an opposite
variation in the gluon 
distribution within the proton. Although this weak dependence on 
$\alpha_s$ defeats the original motivation for studying
$R_{10}$, it turns out, rather surprisingly, that there is a large mismatch 
between the experimental result at the Tevatron and theoretical predictions 
of $W +$ jet(s) production from NLO QCD calculations incorporated in the 
DYRAD Monte Carlo program~\cite{NLO_Wg}. The discrepancy is well above 
three or even four standard deviations for most of the kinematic range 
studied (except 
in the region of very large transverse momentum of the jet where the errors 
are large). A difference of this magnitude between the theoretical 
prediction and 
the experimental data cannot be accounted for by variation of $\alpha_s$ 
and may thus be said to constitute an `$R_{10}$-anomaly'. It is only fair 
to mention, however, that similar studies reported recently by the CDF 
Collaboration~\cite{CDF_R10} are consistent with 
QCD predictions and show no signs of anomalous behaviour.
 
While so large a discrepancy has been reported only recently, a
modest excess in the experimentally measured $R_{10}$ over the 
theoretical prediction has been consistently reported by the D0 
Collaboration for some time~\cite{D0_Wg_old}. Earlier errors being large, 
however, this 
excess (which was at the level of two standard deviations or less) was of
marginal significance and an explanation could perhaps be given in terms 
of the uncertainty in the theoretical predictions 
resulting from the lack of precise knowledge of the gluon flux.
An additional complication arises because gluon distributions 
are not the only source of 
theoretical uncertainty in the QCD predictions --- both the 
experimental determination and the theoretical prediction of $R_{10}$
depend on the proper definition of a jet via some `standard' algorithm. 
This immediately makes the result
susceptible to effects associated with the soft physics 
of jets. Resummation of the soft radiation could possibly lead to 
significant effects --- an issue addressed recently by 
Balazs and Yuan~\cite{BaYu_97}, who present an estimate based on a 
Collins-Soper-Sterman-type~\cite{CoSoSt_85}
 resummation calculation. Unfortunately, the magnitude of 
this effect is small even when compared to the earlier 2$\sigma$ 
disagreement~\cite{CDF_R10,D0_Wg_old} and cannot be held to explain the 
current large excess~\cite{D0_Wg_new}. Another feature of their calculation 
is that the effects of soft gluon radiation, not surprisingly, are 
dependent on the jet transverse energy $E_T$, and tend to be more pronounced 
at the smaller end of the $E_T$ spectrum. The {\it observed} 
discrepancy, however,
is equally significant over the entire range of $E_T^{\rm min}$, the minimum 
required transverse energy of the jet,  (see Fig. 1). 
The highest $E_T^{\rm min}$ 
bins have larger errors, of course, but even in this region the discrepancy 
is substantial and it is rather improbable that the effect at such high 
values of $E_T$ should owe its existence to soft gluon radiation.
 
In view of the above remarks it is difficult to see how the 
`$R_{10}$-anomaly' can be explained in the SM and one is thus tempted to 
consider it an indication for new physics beyond the SM. This is 
strengthened by the obvious fact that the Tevatron probes a 
hitherto-unexplored kinematic region, one
not accessible to the CERN $Sp\bar pS$ collider. One candidate that 
immediately suggests itself \cite{GOUN,W8_pheno,BaurStreng}
is the colour-octet analogue of the $W^\pm$ 
boson, which we denote $W_8^\pm$. Such objects are predicted in a whole 
class of composite models, wherein the  known 
fermions and bosons are assumed to be composed of more 
elementary constituents \cite{PatiSalam} called preons. 
Consider, for example, the {\it haplon} model of Fritzsch and 
Mandelbaum~\cite{HAPLON}, 
which has both spin-$\frac{1}{2}$ and scalar preons transforming 
under the gauge group $SU(3)_c \times U(1)_{em} \times G_H$, 
where $G_H$ ($\equiv U(1), SU(N)$) is a local `hypercolour' symmetry.
Isospin is no longer a local gauge symmetry and the weak interaction 
is interpreted as a residual van der Waals-type force. The matter 
content is given by two spinors $\alpha (3, -\frac{1}{2}, N)$ and 
$\beta (3,\frac{1}{2}, N)$ and two scalars $x (3, -\frac{1}{6}, \bar N)$
and $y (\bar 3, \frac{1}{2}, \bar N)$. The SM particles are composed of two 
preons each and are held together by the hypergluons. Specifically,
the $W^+$ is now nothing but the colour 
singlet\footnote{It is argued that the corresponding 
                 lightest scalar states are heavier~\protect\cite{HAPLON}.}
$\bar{\alpha} \beta$. Clearly, a
colour-octet partner of the $W^+$  naturally exists in this model.

The haplon model is, of course, only one example of a preonic model. Even in
the simplest version of the Pati-Salam model~\cite{PatiSalam}, 
a $W_8$ exists as a bound state of two scalar `chromons'. 
Similarly the {\it rishon} 
models~\cite{RISHON} also predict a $W_8$. A complete survey 
of these models may be found in Ref.~\cite{LYONS}. In this letter, we do
not attempt to study the specific properties of preonic models in detail. 
We simply invoke these to point out that there {\it exist} well-known 
composite models with (possibly light) colour-octet analogues of the 
$W$ boson. Our chief concern is to study the implications of such a particle
for measurements leading to the ratio $R_{10}$. With this rather modest 
aim in view, it is sufficient to parametrize the effective couplings of a 
$W_8$  following Gounaris and Nicolaidis \cite{GOUN}:
\be
\barr{rcl}
{\cal L}_1 &=& \dis 
         - \frac{g_8}{\sqrt{2} }
         {\cal W}_{\mu a}^{+} \; 
              \bar u_L \gamma^\mu \frac{\lambda_a}{2} d_L
      + {\rm h.c.}
     \\[2ex]
{\cal L}_2 & = & 
      - g_B W_\mu^{+} {\cal W}_{\nu a}^{-}G^{\mu\nu}_{a}
      - g'_B \epsilon^{\mu\nu\lambda\sigma}
          W_\mu^{+} {\cal W}_{\nu a}^{-} G_{\lambda\sigma a}
      + {\rm h.c.}
\earr                                           
      \label{lagrangian}
\ee
where $a=1,\dots, 8$ 
and $G^{\lambda\sigma a}$ represents the gluon field-strength tensor.
The term ${\cal L}_2$ represents the interaction of the $W_8$ with a $W$ and
a gluon via a parity-conserving term with coupling $g_B$ 
and a parity-violating term with coupling $g_B^{\prime}$ respectively.
Since these couplings arise in effective interactions,
it is also of interest to consider how they scale with increase
in the centre-of-mass energy of the process under consideration, especially
as one approaches the scale of the underlying new physics. To
account for such effects without calculating them in a
specific model, we assume that the coupling constants scale as
\begin{equation}
g_{8} (Q^2) = g_{8} (0) \; \left[ 1 + \frac{Q^2}{\Lambda^2} \right]^{-n}
            \label{formfac}
\end{equation}
and similarly for $g_B, g_B^\prime$,
where the scaling index $n = 0,1,2,...$ and $\Lambda$ is
the compositeness scale (${\cal O}(1) \tev$).

We can now calculate various processes involving the $W_8$. Within 
the SM, a $W +$ jet final state can occur due to 
either $q \bar q' \rightarrow W g$ or $q g \rightarrow q' W$. 
The presence of $W_8$ simply introduces an $s$-channel diagram in the
first case and a $t$-channel one in the second. 
The parton-level cross-section for the process $u \bar d \rightarrow W^+ g$
can be expressed as 
\be
\barr{rcl}
\dis
\frac{ {\rm d} \hat \sigma }
     { {\rm d}\th }       (u \bar d \rightarrow W^+ g)
& = & \dis
\frac{1}{16 \pi \sh^2} 
         \left[ T_{SM} + T_{W8} + T_{int} \right]
             \\[2.0ex]
T_{SM} & = & \dis
    \frac{2 g^2 g_s^2}{9} 
\left[ \frac{\th^2 + \uh^2 + 2 M_W^2 \sh}{\th\uh} \right] 
             \\[2.5ex]
T_{W8} & = & \dis
             \frac{g_8^2 (g_B^2 + 4g_B^{\prime 2} ) }
                  {9M_W^2}
         \left[ \frac{ (\sh +2 M_W^2)(\th^2 +\uh^2)}
                     {(\sh-M_8^2)^2+(M_8\Gamma_8)^2} \right] 
              \\[2.5ex]
T_{int} & = & \dis
          -\frac{4 g g_s g_8 }{9} 
          \left[ \frac{g_B M_8 \Gamma_8 (\sh - M_W^2) 
                          - 2 g_B^\prime (\sh + M_W^2) (\sh - M_8^2) }
                      { (\sh - M_8^2)^2 + (M_8 \Gamma_8)^2}
\right] \ ,
                   \label{matelsq}
\earr
\ee
with the decay width $\Gamma_8$ given by
\be
\barr{rcl}
\Gamma_8 & = & \dis\sum_i \Gamma(W_8 \rightarrow u_i\bar d_i) 
         + \Gamma(W_8 \rightarrow Wg) 
         \\[2ex]
\Gamma(W_8 \rightarrow u\bar d) & = & \dis
         \frac{g_8^2 M_8}{192 \pi} 
         \; \left[ 2 - (x_u + x_d) -  (x_u - x_d)^2
            \right] 
         \; \lambda(1, x_u, x_d)
         \\[2ex]
 \Gamma(W_8 \rightarrow Wg) & = & \dis
         \frac{g_B^2 + 4 g_B^{\prime 2} }{96 \pi} \; M_8
         \; (1 - x_W)^3 \left( 1 + \frac{1}{x_W} \right) \ ,
\earr
         \label{width}
\ee
where $\lambda(a,b,c) \equiv \sqrt{ (a - b - c)^2 - 4 b c}$ and 
$x_i \equiv m_i^2 / M_8^2$. 
The LO cross-section(s) for the process(es) $qg \rightarrow q' W^+$ can 
be obtained from eq.(\ref{matelsq}) simply by exploiting the 
crossing symmetry. 

In order to obtain the hadron-level cross-sections,
these formulae need to be convoluted with the corresponding 
parton  densities in the incoming proton-antiproton pair. We 
do this by using the CTEQ3M~\cite{CTEQ} distributions, which were 
calculated using the package {\sc pdflib}~\cite{PDFLIB}. 

To identify the $W$, the D0 experiment has used the $e\nu$ decay 
channnel~\cite{D0_Wg_new,D0_Wg_old}. The final state thus consists of a
hard electron accompanied by large missing energy and one or more jets.
The angular coverage and the energy threshold of the hadronic calorimeter
requires that 
\be
       E_{\rm jet} \geq 20 \gev, \qquad   
        | \eta_{\rm jet} | < 4  \ ,
                           \label{jet_cut}
\ee
for a jet to be identified as one. Here $\eta$ refers to the 
pseudorapidity. For the purpose of this measurement, the experiment 
concentrated on relatively central but hard electrons with 
\be
       p_T (e) \geq 25 \gev, \qquad   
        | \eta_e | < 1.1                    \ .
                           \label{e_cut}
\ee
Further, the electron was required to be isolated from the jet(s)
by imposing a fixed cone algorithm with angular separation between the 
electron and any jet
\be
         \Delta R \equiv \sqrt{ (\delta \eta)^2 + (\delta \phi)^2}
                  > 0.4         
              \ ,
                           \label{delta_r}
\ee
where $\delta \phi$ is the difference in the azimuthal angles. 
Events
with more than one electron track passing the above selection criteria were
removed to eliminate background from $Z$ decays. And finally, an event 
was required to have a minimum missing transverse momentum:
\be 
    \ptsl \: > 25 \gev
             \ .
       \label{pT_cut}
\ee

\begin{figure}[h]
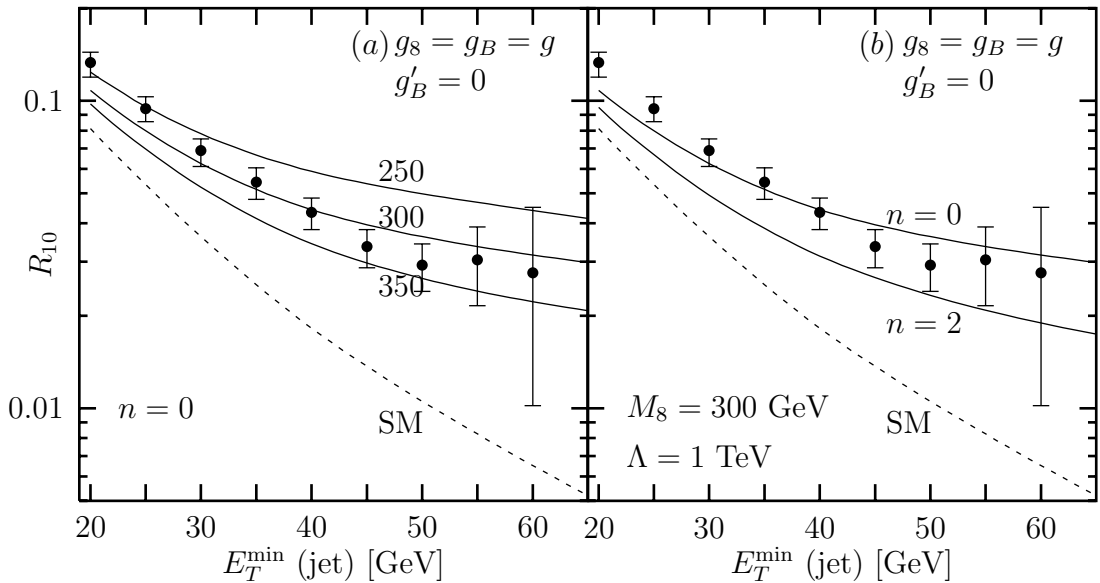

\begin{center}
\vspace*{-10pt}\hspace*{14pt}      
                    \input{fig_1a.psl}
\vspace*{-144pt}\hspace*{-90.5pt} 
                    \input{fig_1b.psl}
\end{center}
\vspace*{130pt}
         \caption{\em Ratio of $W + 1$ jet to $W + 0$ jet production 
           cross-sections at the Tevatron in the presence of a $W_8$ with 
           $g'_B = 0$. $E_T^{min}$ (jet) represents the cut on minimum
           transverse energy of the jet. The SM plus $W_8$ (LO) results
           are shown for {\em (a)} three illustrative values of $M_8$ 
           (marked, in GeV) and scale-independent couplings $g_B$ and 
           $g_8$; {\em (b)} $M_8 = 300 \gev$ and form-factor
           exponents  (Eq.\protect\ref{formfac}) $n = 0,2$. 
           The form-factor is assumed to be 
           identical for both $g_B$ and $g_8$,
           and the compositeness scale is taken as $\Lambda = 1$ TeV.}
    \label{fig:g_B}         
\end{figure}

Using a set of cuts closely modelled on the above, we make a parton-level
Monte Carlo estimate of the value of 
$R_{10}$ in the presence of the $W_8$. For this purpose, we 
use the NLO results for the SM contribution, but only the LO
results of Eq.(\ref{matelsq}) for the additional contribution 
due to the $W_8$. The numerical results are presented in Fig.\ref{fig:g_B}
As a guide we refer once again to the haplon model \cite{HAPLON} and 
note that it predicts $M_8 - M_W \sim \alpha_s \Lambda$ which 
immediately leads to $M_8 \sim 200$--300 GeV. 

Now, a change in $E_T^{\rm min}$, the minimum hadronic transverse 
energy required of an event, is expected to lead to a strong 
variation in $R_{10}$. This is especially true of the 
SM contribution as the radiated jets are predominantly soft. 
We note that the $W+0$ jet cross-section receives a substantial
contribution from $W+1$ (or more) jet events where the jet fails to 
satisfy the selection criteria. 

In Fig. \ref{fig:g_B}, we illustrate the effect of a 
$W_8$ with typical values of the mass $M_8 = 250$, 300 and
350 GeV. 
These are shown by solid lines, to be compared with the
SM prediction (dotted line) and the D0 data with $1\sigma$ error bars. 
For this figure, we assume the couplings to have no scale-dependence,
{\em i.e.} $n = 0$, and that there is no parity violation in the 
$gWW_8$ coupling, {\em i.e.} $g'_B = 0$. It may be observed that
we now have a fairly good agreement with the data within the errors
for $M_8 = 300$~GeV, while the other values lead to cross-sections which 
are too large or too small, as the case may be. However, this is not
very restrictive, since the cross-sections scale as the couplings
$g_8,g_B$. Figure \ref{fig:g_B}$b$\  illustrates the effect of considering
a form-factor-like behaviour of the couplings with $n = 2$ while the 
$n = 0$ case is also shown for comparison. Clearly, for $n = 2$,
$M_8 = 300$~GeV is no longer a suitable choice; $M_8 \sim 250$~GeV
appears to be a better choice. With the present state of our knowledge 
of the masses, couplings and scale-dependence, however, it is hard to
be more specific than to say that couplings of electroweak strength
and $M_8 = 300 \pm 100$~GeV might explain the observed discrepancy.

\begin{figure}[h]
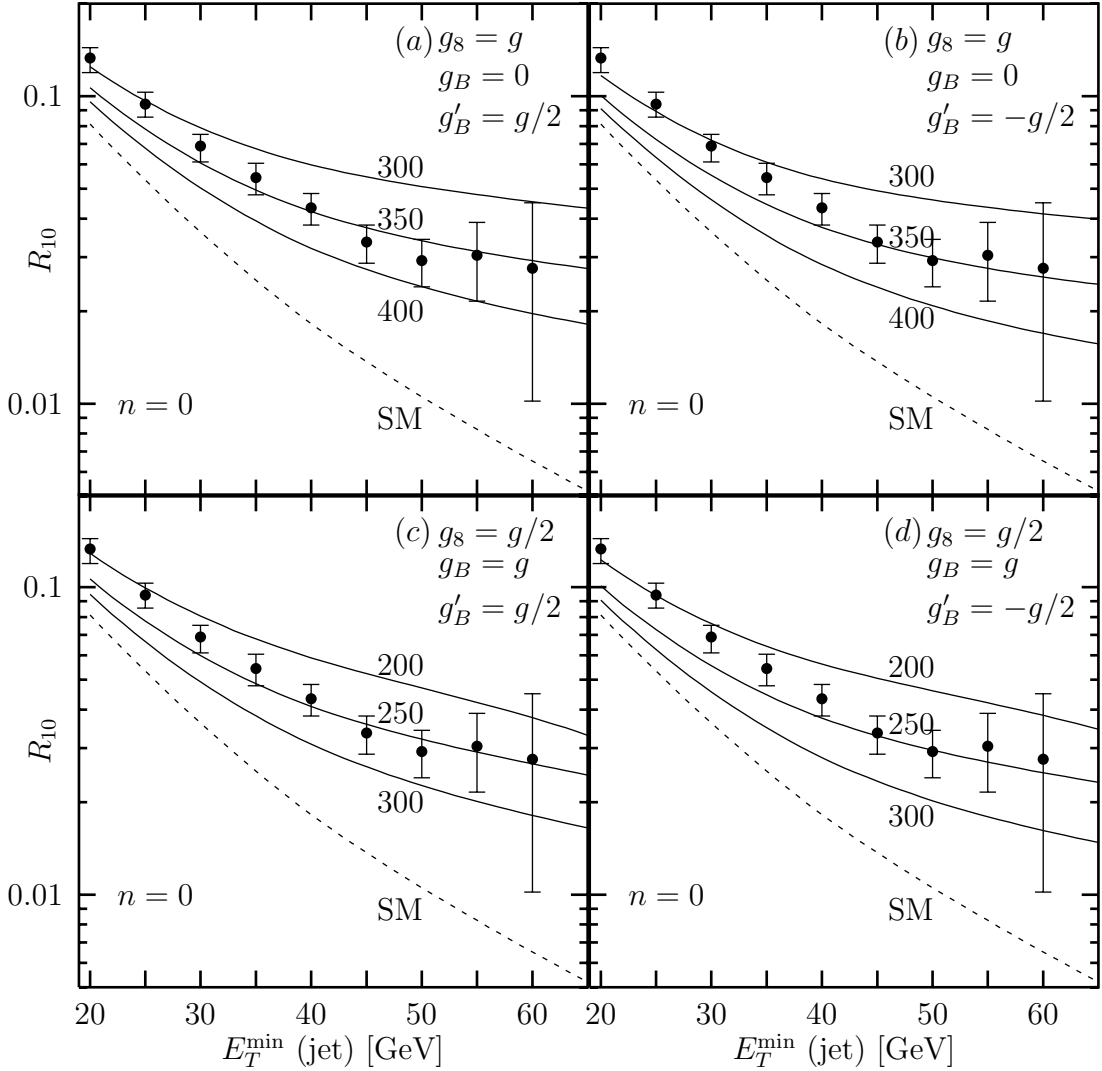

\begin{center}
\vspace*{-3pt}\hspace*{14pt}      
                    \input{fig_2a.psl}
\vspace*{-144pt}\hspace*{-89.5pt} 
                    \input{fig_2b.psl}

\vspace*{112pt} \hspace*{10pt}      
                    \input{fig_2c.psl}
\vspace*{-144pt}\hspace*{-89.5pt} 
                    \input{fig_2d.psl}
\end{center}
\vspace*{130pt}
         \caption{\em As in Fig.\protect\ref{fig:g_B} but 
           for various combinations of $g_B$ and $g'_B$. 
           Couplings are assumed to be scale-invariant.}
    \label{fig:g_B&g_Bp}         
\end{figure}

A closer look at Fig. \ref{fig:g_B} will reveal that although the 
relevant solid curve fits the data within the errors, there is still
room for improvement. This is because the curve is somewhat flatter
than the general trend of the data seems to indicate. One possible
remedy might be to consider a non-vanishing $g'_B$, which means 
a parity-violating $gWW_8$ interaction. In Fig. \ref{fig:g_B&g_Bp} 
we have illustrated the effect of this for the combinations
\[
\barr{rlcrl}
{\rm (a)} & g_8 = g, \;   g_B = 0,  \; g'_B =  g/2, 
     & \qquad & 
    {\rm (b)} & g_8 = g, \;   g_B = 0,  \; g'_B = -g/2, 
       \\
{\rm (c)} & g_8 = g/2,  \; g_B = g,  \; g'_B =  g/2, 
     & \qquad &
{\rm (d)} & g_8 = g/2,  \; g_B = g,  \; g'_B = -g/2, 
\earr
\]
and for suitably chosen values of the mass $M_8$ in the range 
200--400 GeV. A close look will reveal that the shape of the curve(s)
changes and now resembles the trend of the data more accurately,
%
%
although it is difficult to find a combination of masses and
couplings that will fit every single data point within $1\sigma$
without resorting to rather fine tuning. At the present level of
accuracy, this may not even be desirable, so we can only conclude
that the data may contain a hint of non-vanishing $g'_B$ interactions.
It is also noteworthy
that higher-order corrections to the $W_8$-induced diagrams
will tend to change the shape of the curves and, in particular, it
is plausible that soft-gluon resummation effects in the $W_8$-induced
diagrams will tend to make the low-$E_T$ part of the spectrum steeper
than the behaviour shown by the LO results presented here. The question
of $g'_B$ interactions remains, therefore, an open one.

As the above comments show, a colour-octet $W_8$ boson with requisite
masses and couplings may be the answer to the excess in $W+1$ jet events
seen by the D0 Collaboration at the Tevatron. We must now consider the 
possibility that a new interaction of this nature may have observable
consequences in other experiments. Perhaps the most obvious ones to 
suggest are low-energy measurements of flavour-changing neutral current
(FCNC) processes such as neutral meson mixing and rare $K$ and $B$ meson
decays. These would be affected if the current in Eq.(\ref{lagrangian}) 
contained two quark fields of different flavours
with a mixing angle factor, as indeed seems quite natural since the 
colour-singlet current must have this feature. Even if we assume that 
the mixing elements in the octet sector are of the same size as those
in the singlet sector, it is easy to see that the additional contributions
to $\Delta m_K$ or $\Delta m_B$ are well below the SM values for the 
parameter ranges of interest. 
Furthermore, ($a$) the mixing angles appearing in the colour-octet $W_8$ 
interactions need not always be identical with the Cabbibo-Kobayashi-Maskawa 
matrix \cite{BaurStreng} and ($b$) there could be other contributions to FCNC
amplitudes from exotics such as colour sextet quarks. Thus, 
FCNC processes are unlikely to yield definitive constraints on 
$W_8$ interactions and we do not consider them any further.

We also need to consider possible constraints from 
the electroweak precision variables
as measured at LEP. The decay $Z \rightarrow b \bar b$
giving rise to the well-measured parameter $R_b$ 
is the simplest process to be affected through vertex
corrections involving a $W_8$. These do not require flavour-changing
vertices and hence cannot be wished away. If the only interactions given 
in the theory are those given by Eq.(\ref{lagrangian}), then, indeed,
one can obtain fairly stringent constraints on the coupling $g_8$. 
However, great caution needs to be exercised in such an attempt 
since Eq.(\ref{lagrangian}) 
is evidently not the whole story. By itself, it would give a divergent 
contribution to $R_b$ and both the other relevant interactions 
(\eg, $Z W_8 W_8$) and particles (\eg, $Z_8$) in the model 
have to be considered for a definitive statement to be 
made. Similar arguments also hold for the 
$\rho$-parameter\footnote{Direct preonic contributions 
        can also be relevant, although the numbers depend 
        rather strongly on the preon dynamics~\protect\cite{Baur_84}.}.
With so many free parameters (although they are 
calculable in principle, assuming one can handle the dynamics of
preonic interactions) and mutually cancelling contributions, 
it seems a phenomenologically sound procedure to 
ignore the $R_b$ constraint altogether, and this is what has been
done in the present study. It must be pointed out, though, that 
the naive bounds may be evaded by simply scaling down $g_8$ and 
compensating the effect by scaling up $g_B$ ($g'_B$). 

It might seem, from the above discussion, that models with colour-octet
$W_8$ bosons are difficult to pin down because of a plethora of 
unknown particles and parameters which can be varied at will.
Does this mean that the $W_8$ solution of the $R_{10}$ anomaly
loses the prime virtue of falsifiability? Fortunately, the answer is,
No. At the Tevatron itself, one can check for processes involving 
the $g_8$ and $g_B(g'_B)$ couplings {\em separately} (and no others). 
In principle,
non-observation of both would rule out the solution suggested in
this letter since, as explained above, one can tolerate a reduction
in one or the other, but not in both. We thus turn to a discussion
of other observables at the Tevatron.

In $p\bar p$ collisions at 1.8 TeV, 
a 250--300 GeV $W_8$ would naturally contribute 
to processes of the 
form $q_1 \bar q'_2 \rightarrow q_3 \bar q'_4$ resulting in a possible 
enhancement of dijet rates at the Tevatron~\cite{CDF_dijet, CDF_dijet_new}. 
As in the case of $W + 1$ jet production, the leading $W_8$ contribution 
arises from the resonance diagram, and hence, it is instructive to
concentrate on it. Once produced, the $W_8$ can decay into either 
of dijet or the $gW$ channel, with the relative branching fractions 
given by Eq.(\ref{width}). To appreciate the 
CDF bounds on new particles decaying into dijets~\cite{CDF_dijet_new}, 
let us compare the production rate for $W_8$ with that of a $W'$ in 
an extended gauge model. It is easy to see that, for identical 
mass and coupling, 
$$\sigma (p \bar p \raisebox{-0.5ex}{$\stackrel{W_8}{\:\longrightarrow\:}$}
              u \bar d + X)  
   = \frac{2}{9}\;
\sigma (p \bar p \raisebox{-0.5ex}{$\stackrel{W'}{\:\longrightarrow\:}$}
               u \bar d + X).
$$
Thus, the enhancement in the dijet rate is well below the constraints of 
Fig.3 of Ref.~\cite{CDF_dijet_new} for $g_8 \approx g$. 
However, as the statistics improve with time, we can expect an improvement 
in this constraint.

\begin{figure}[h]
\begin{center}
                    \input{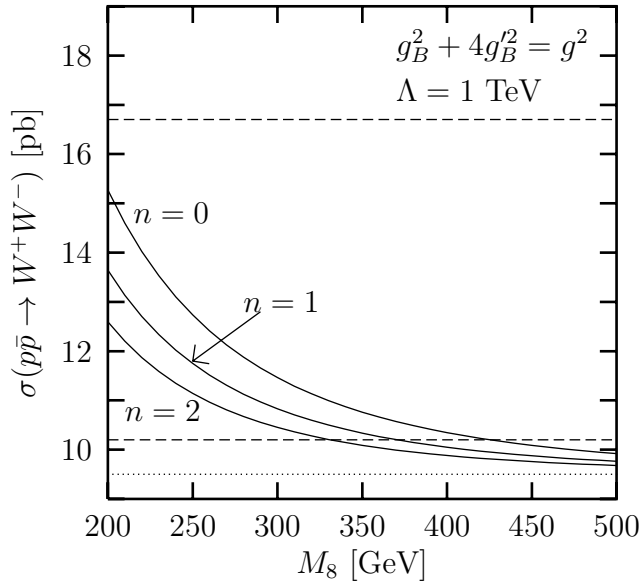}
\end{center}
\vspace*{-10pt}
         \caption{\em $W^+W^-$ production cross-section at the 
           Tevatron as a function of $M_8$. For the $W_8$-mediated
           process, only the LO contribution has been included while 
           the NLO result for the SM has been taken from 
           Ref.\protect\cite{WW_NLO}. The three solid curves 
           correspond to three different values of the form-factor 
           exponent (Eq. \protect\ref{formfac}) $n = 0,1,2$. 
           The dotted line corresponds to the SM value, while 
           dashes give the experimental central value 
           and the $1\sigma$ upper bound from 
           CDF~\protect\cite{CDF_WW}.}
         \label{fig:WW_cross}
 \end{figure}

Another immediate consequence of the Lagrangian of Eq.(\ref{lagrangian}) 
is that the process $g g \rightarrow W^+ W^-$ is now allowed at the 
tree-level through $t$- and $u$-channel exchange of the $W_8$. The 
corresponding differential cross section is given by
\begin{equation}
  \begin{array}{rcl}
   \dis   \frac{{\rm d} {\hat \sigma}} {{\rm d} \th} 
         (g g \rightarrow W^+ W^-)
             & = & \dis \frac{\pi}{\sh^2} \:
                        \left( \frac{ g_B^2 + 4 g_B^{\prime 2} }
                                    {16 M_W^2 M_8^2 }
                        \right)^2
                   \: \left[   \frac{ f(\th)}{ (\th - M_8^2)^2}
                             + \frac{ f(\th,\uh)}
                                    { (\th - M_8^2) (\uh - M_8^2) }
                             + ( \th \leftrightarrow \uh)
                      \right]
               \\[3ex]
         f(\th) &\equiv & 
                  M_8^4 \th^2 ( 2 \sh^2 + 2 \sh \th + \th^2) 
                        - 4 M_W^2 M_8^4 \th (2 \sh^2 + 3 \sh \th + \th^2) 
               \\[1.5ex]
              & & \hspace*{1em} 
                 + 2 M_W^4 \left[\th^2 (4 M_8^2 \sh + \th^2) 
                                   + M_8^4 ( 4 \sh^2 + 9 \sh \th + 3 \th^2) 
                           \right]
               \\[1.5ex]
              & & \hspace*{1em} 
                 - 4 M_W^6 \left[M_8^4 (2 \sh + \th)
                                   + 4 M_8^2 \sh \th + 2 \th^3
                           \right]
               \\[1.5ex]
              & & \hspace*{1em} 
                 + M_W^8 (  M_8^4 + 8 M_8^2 \sh + 12 \th^2)
                 - 8 M_W^{10} \th + 2 M_W^{12}
               \\[2ex]
         f(\th,\uh) &\equiv & 
                  M_8^4 \th \uh (\th^2 + 3 \th \uh + \uh^2)
                    - 4 M_W^2 M_8^4 \th \uh (\th + \uh)
               \\[1.5ex]
              & & \hspace*{1em} 
                  + M_W^4 \left[ \th^2 \uh^2 
                                + M_8^4 \left\{ 4 (\th^2 + \uh^2) + 10 \th \uh
                                       \right\}
                          \right]
               \\[1.5ex]
              & & \hspace*{1em} 
                  + 8 M_W^6 M_8^2 (\th + \uh) (\th + \uh - 2 M_8^2)
               \\[1.5ex] 
              & & \hspace*{1em} 
                  + M_W^8 \left[ 17 M_8^4 + \th^2 + \uh^2 - 8 M_8^2 (\th + \uh)
                          \right]
               \\[1.5ex] 
              & & \hspace*{1em} 
                  - 4 M_W^{10} (\th + \uh)
                  + 5 M_W^{12} \ .
  \end{array}
  \label{ggWW}
\end{equation}
Consequently, one may expect to see deviations in the $W$-pair 
cross-section as measured at the Tevatron from the SM predictions.
In Fig.\ref{fig:WW_cross}, we present the variation of the total
cross section as a function of the $W_8$ mass for 
$g_B^2 + 4 g_B^{\prime 2} = g^2$. For smaller values of $M_8$, the 
deviation is quite significant and this effect could thus serve as 
a discriminant for our explanation of the $W + 1$ jet excess. 
However, in view of the large 
statistical errors in the measurement~\cite{CDF_WW}, we are still 
some way from a definitive statement.

It is interesting to 
note that the cross-section in Eq.(\ref{ggWW}) {\em grows} with the 
energy. This is a reflection of the lack of $SU(2)$ gauge invariance 
in the theory and of the underlying compositeness. 
This energy behaviour can be cured either by postulating 
an energy-dependence of the couplings (see Eq.(\ref{formfac}) and 
Fig. \ref{fig:WW_cross}) or through the introduction of a $Z_8$ with 
the right couplings.

To summarise, then, the reported discrepancy between the experimental
determination of the ratio $R_{10}$ and the SM prediction seems
to be of a magnitude not easily explicable in terms of conventional effects
such as soft-gluon resummation. If the discrepancy persists even after
more careful analyses have been done and more data are available,
then it is very likely to be due to some new physics. 
The fact
that this effect shows up at energies close to the electroweak scale 
suggests that the mechanism of electroweak symmetry-breaking may be 
intimately linked to
the physics of strong interactions. Such an interpretation is
also viable in the context of the recently discovered large-$Q^2$ 
anomaly at HERA \cite{hera}, and has been investigated in a number of
recent theoretical papers \cite{heraex}. 
It is interesting to note that the haplon
model, among others, predicts a light leptoquark state $(\bar x y)$ 
with quantum numbers $(3,\frac{2}{3},1)$ which will decay into either
of an $e^+d$ or a $\nu u$ final state --- this is just the kind of 
new particle that seems to be indicated by the HERA high-$Q^2$ anomaly.
While the HERA issue is still a debatable one, 
it is undoubtedly true that in some of the 
well-known composite models, colour-octet incarnations of the $W$
boson are predicted; we find that the inclusion
of these coloured bosons through the effective interactions
given above helps resolve the discrepancy between the data and
theory. The mass of the $W_8$ required for an agreement with the
data is in excess of 250~GeV, which might explain why 
its effects have not yet been 
seen in other experiments. The results presented in this paper
suggest that the time is ripe to carry out a detailed, global
analysis of the data from the LEP and the Tevatron to study if there
are other manifestations of this kind of new physics. It is just
possible that we are standing at the threshold of a new era of
sub-quark and sub-lepton physics which is just beginning to show up
in deviations from the SM at the edge of the kinematic range studied
till now.  We can thus look forward to an exciting period as
as more data are acquired and analysed at the running high-energy 
experiments.

DC and SR would like to acknowledge helpful discussions with J.C. Pati.

\end{document}